# An Interim Summary on Semantic Model Differencing [*]

Shahar Maoz
School of Computer Science
Tel Aviv University, Israel

Jan Oliver Ringert, Bernhard Rumpe
Software Engineering
RWTH Aachen University, Germany

## ABSTRACT

This position paper provides an interim summary on the goals and current state of our ongoing research project on semantic model differencing for software evolution. We describe the basics of semantic model differencing, give two examples from our recent work, and discuss future challenges in taking full advantage of the potential of semantic differencing techniques in the context of models' evolution.

## Categories and Subject Descriptors

D.2.2 [**Software Engineering**]: Design Tools and Techniques; D.2.4 [**Software Engineering**]: Software/Program Verification

## 1. INTRODUCTION

In a recent workshop paper [8] we have presented our vision on semantic model differencing.

Most existing approaches to model differencing concentrate on matching between model elements using different heuristics related to their names and structure and on finding and presenting differences at a concrete or abstract syntactic level. While showing some success, these approaches are also limited. Models that are syntactically very similar may induce very different semantics (in the sense of 'meaning' [3]), and vice versa, models that semantically describe the same system may have rather different syntactic representations. Thus, a list of syntactic differences, although accurate and complete, may not be able to reveal the real implications these differences have on the correctness and potential use of the models involved. In other words, such a list, although easy to follow, understand, and manipulate (e.g., for merging), may not be able to expose and represent the semantic differences between two versions of a model, in terms of the bugs that were fixed or the features (or new bugs...) that were added.

[*]J.O. Ringert is supported by the DFG GK/1298 AlgoSyn.

We are interested in **semantic model differencing**. Specifically, our vision is to develop **semantic diff operators** for model comparisons: operators whose input consists of two models and whose output is a set of **diff witnesses**, instances of the first model that are not instances of the second. Such diff witnesses serve as concrete proofs for the real change between one version and another and its effect on the meaning of the models involved. In particular refactorings, which change the internal structure or even only the presentation of a model, but do not change the semantics, can be identified.

## 2. BASIC DEFINITIONS

The fundamental building block of our approach is an abstract semantic differencing operator that accepts two models as input and outputs a set of diff witnesses, each of which is an instance in the semantics of the first model and not in the semantics of the second.

More formally, we consider a modeling language $ML = \langle Syn, Sem, sem \rangle$ where $Syn$ is the set of all syntactically correct expressions (models) according to some syntax definition, $Sem$ is a semantic domain, and $sem : Syn \rightarrow \mathcal{P}(Sem)$ is a function mapping each expression $e \in Syn$ to a set of elements from $Sem$ (see [3]).

The semantic diff operator $\mathit{diff} : Syn \times Syn \rightarrow \mathcal{P}(Sem)$ maps two syntactically correct expressions $e_1$ and $e_2$ to the (possibly infinite) set of all $s \in Sem$ that are in the semantics of $e_1$ and not in the semantics of $e_2$. Formally:

DEFINITION 1. $\mathit{diff}(e_1, e_2) = \{s \in Sem |\ s \in sem(e_1) \land s \notin sem(e_2)\}$.

Note that $\mathit{diff}$ is not symmetric, $\mathit{diff}(e_1, e_1) = \emptyset$, and $\mathit{diff}(e_1, e_2) \cap \mathit{diff}(e_2, e_1) = \emptyset$. The elements in $\mathit{diff}(e_1, e_2)$ are called *diff witnesses*.

The above definition can be used to induce an abstract set of relations between models that is of interest in high-level evolution analysis. Specifically, when considering two models, one model may be a refinement of the other (has a restricted semantics), an abstraction of the other (has a more permissive semantics), equivalent to the other (has equal semantics), or incomparable to the other (each model's semantics allows instances the other does not allow). When applied to the version history of a certain model, these relations provide a high-level semantic insight into the model's evolution, which is not available in existing syntactic approaches.

An important characteristic of our approach is that we do not look for a succinct mathematical representation of all differences between the two models. Rather, we believe that in order to make semantic differencing useful and attractive to engineers, one needs to take a constructive and concrete approach: to compute and present concrete, specific, and thus easy to understand witnesses for the difference.

The abstract semantic differencing operator *diff* is intentionally language independent and does not relate to the question of how semantic differences should be computed. This allows us to investigate properties that are common to semantic evolution analysis in general. Concrete differencing operators for specific languages, and related techniques (e.g., abstraction, summarization) are addressed separately.

## 3. WORK TO DATE

To date, we have developed two semantic differencing operators, one for class diagrams and one for activity diagrams. We describe these briefly below.

### 3.1 CDDiff

Class diagrams (CDs) are widely used for modeling the structure of object-oriented systems. The syntax of CDs includes classes and the various relationships between them (association, generalization, etc.). We consider the semantics of a class diagram to consist of the (possibly infinite) set of object models it allows.

In [7] we have presented CDDiff, a semantic differencing operator for class diagrams. CDDiff takes two class diagrams as input and outputs a set of diff witnesses, each of which is an object model that is an instance of the first class diagram and not of the second. CDDiff is computed using a reduction to Alloy [4].

### 3.2 ADDiff

Activity diagrams (ADs) have recently become widely used in the modeling of work flows, business processes, and web-services, where they serve various purposes, from documentation, requirement definitions, and test case specifications, to simulation and code generation. We consider an operational semantics for activity diagrams (as presented in [6]), which induces a trace-based semantics: the semantics of an activity diagram consists of the set of execution traces of actions it allows.

In [5] we have presented ADDiff, a semantic differencing operator for activity diagrams. ADDiff takes two activity diagrams as input and outputs a set of diff traces, each of which is an execution trace that is possible in the first activity diagram and not in the second. ADDiff is computed using a symbolic fixpoint algorithm, inspired by symbolic model-checking [2], using Binary Decision Diagrams (BDDs), and is implemented using the APIs of JTLV [10].

## 4. FUTURE WORK CHALLENGES

We list future work challenges in the semantic differencing project.

**Additional language-specific differencing operators** In contrast to syntactic differencing, semantic differencing is language-specific: the concrete definitions and the methods used to determine equality or compute diff witnesses and to present the results are unique to each language. To date we have defined two concrete semantic differencing operators, for class diagrams and for activity diagrams. To further extend the applicability of semantic differencing, one needs to define differencing operators for additional languages, covering structural and behavioral modeling (e.g., statecharts), as well as variability modeling (feature models), transformation modeling (various model transformation languages), and various domain specific modeling languages.

**Abstraction** Abstraction, a fundamental concept in model-driven engineering, has an important role in the context of model comparisons. Specifically, two models may be equivalent at one level of abstraction but different in a less abstract level. Thus, the level of abstraction of interest should be defined by the engineer applying the comparison, who may be aware that the models at hand differ at a certain detailed level, but would be interested in comparing them at a higher level, where they are supposedly equivalent.

The motivation for defining abstractions for model comparisons is twofold. First, computational complexity and performance. Abstract models are smaller and simpler and thus, their comparison may be faster. Second, engineer interest. Based on the task at hand and on external knowledge about the models involved, the engineer may be interested in restricting the comparison to only some aspects of the models, while ignoring others.

As an example for the use of abstractions in model comparison, consider attribute abstraction in CDs comparison. With this abstraction in effect, the operator ignores differences that are caused only by local changes to the attribute lists of the classes in the diagrams. That is, all class attributes of primitive or library types are abstracted away, so that two CDs whose sole difference is at the attributes level are considered equivalent. The attribute abstraction becomes useful when the engineer is aware of attribute-level differences resulting from local changes, but is interested in checking for more global semantic differences, if any. Another application of this abstraction relates to performance. Given two large CDs, with many classes or many attributes per class, one can start by a comparison with the abstraction in effect. If a difference is found, indeed this proves that the CDs' semantics are different. If a difference is not found, however, one has no choice but to make the comparison again without the abstraction. This example has appeared in [7].

**Summarization** In many cases the complete set of diff witnesses is too large to be efficiently computed and effectively presented. Moreover, many of the witnesses are very similar and hence not interesting. Thus, an important challenge of our semantic differencing approach relates to witnesses selection.

To address this challenge we propose using a summarization technique, based on a notion of equivalence that partitions the set of diff witnesses. We envision the result of the computation to be a summarized set, consisting of a single representative witness from each equivalence class. Moreover, we want the computation to be efficient and to not require the enumeration of all witnesses.

As an example, consider the ADDiff semantic differencing operator we have presented in [5]. Due to partial order between forked branches, the number of different action traces that are possible in the first activity diagram and not in the second may be exponential in the size of the input diagrams. Moreover, many of these traces could be very similar and thus not interesting. An equivalence relation that considers two traces that include the same set of actions but in different order equivalent may induce a useful summarization technique for this case.

An alternative approach to summarization could have been the computation of a single, abstract symbolic representation of all differences, as a succinct representation of the differencing operator result. For example, when comparing two state machines models, one may suggest to compute a state machine that accepts exactly all traces that are possible in the first state machine and not in the other. However, we are not looking for such abstract representations. Rather, we intentionally look for concrete, specific instances that are easy to be understood by engineers. Thus, the need for summarization using representatives from equivalence classes rather than using some single symbolic presentation.

Preliminary work we have already done on summarizing semantic differences was presented in [9].

**Presentation** We consider the effective presentation of diff witnesses to the engineer to be a critical part of the potential of semantic differencing to improve evolution analysis.

Presentation of diff witnesses may be textual or visual. Moreover, we expect it to be language specific. For example, for class diagram differencing, differencing object models may be visually presented using generated object diagrams; for activity diagrams differencing, differencing execution traces may be visually presented on the input activity diagram themselves, e.g., by coloring and numbering the nodes that participate in the differencing trace on both diagrams, from the initial node up until the point where the two diagrams differ. Alternatively, one may use a collaboration diagram like notation, possibly with the aid of animation.

One important challenge of presenting differencing witnesses relates to context switching. In the examples mentioned above, class diagram differencing presents object diagrams and thus require cognitive context switching. In contrast, activity diagram differencing presents the differencing traces on top of the diagrams themselves and thus does not require context switching. We consider avoiding context switching to be a major advantage. Brosch et al. [1] have described a technique that avoids context switching in presenting conflicts between model versions. We may be able to develop similar techniques that will be built around the presentation of semantic differences.

Finally, another technique that may be used is animation. We believe that animation may assist in presenting dynamic results, specifically in the context of behavior models or to demonstrate semantic differences over time.

**Integration with syntactic differencing** Many works have suggested various syntactic approaches to model differencing. These works compare the concrete or the abstract syntax of two models and output a set of edit operations (additions, deletions, renames) representing syntactic difference.

We envision the development of techniques that will integrate existing syntactic differencing approaches with our new semantic differencing operators. We give two examples for the potential contribution of this kind of integration.

- Extension of the applicability of semantic differencing in comparing models whose elements have been renamed or moved in the course of evolution, by applying a syntactic matching before running a semantic differencing. Matching renamed model elements (in possible structured namespaces) is a very important problem in practice and many researchers have suggested heuristics to address it. The integrated solution would result in a mapping plus a set of diff witnesses.

- Usage of information extracted from syntactic differencing as a means to localize and thus improve the performance of semantic differencing computations. As a concrete example, identifying syntactically identical sub models is, in many cases, computationally far less expensive than identifying semantically equivalent sub models. Under appropriate assumptions (dependent on the models' context, the specific modeling language used etc.), identical sub models imply identical semantics thus this knowledge may be used to improve the performance of semantic differencing computations.

Note that some of these integrated techniques may be general but others are expected to be language specific.